\documentclass[a4paper,11pt]{article}
\pdfoutput=1
\usepackage[T1]{fontenc}
\usepackage{jcappub}
\usepackage{graphics}
\usepackage{amsmath}
\usepackage{amssymb}
\usepackage{soul}
\usepackage{aas_macros}
\newcommand{\be}{\begin{eqnarray}}
\newcommand{\ee}{\end{eqnarray}}

\renewcommand{\d}{\mbox{${\rm d}$}}

\RequirePackage[normalem]{ulem} %DIF PREAMBLE
\RequirePackage{color}\definecolor{RED}{rgb}{1,0,0}\definecolor{BLUE}{rgb}{0,0,1} %DIF PREAMBLE
\title{Gravitational antiscreening in stellar interiors}
\author[a,b]{A. Bonanno}
\author[c,d]{R. Casadio}
\author[e]{A. Platania}

\affiliation[a]{INAF, Osservatorio Astrofisico di Catania, via S. Sofia 78, I-95123 Catania, Italy}
\affiliation[b]{INFN,  Sezione di Catania,  via S. Sofia 64, I-95123, Catania, Italy.}
\affiliation[c]{Dipartimento di Fisica e Astronomia, Universit\'a di Bologna, Via Irnerio-46, 40126 Bologna, Italy}
\affiliation[d]{INFN, Sezione di Bologna, I.S.~FLAG, 
Viale Berti-Pichat 9/2, 40127 Bologna, Italy}
\affiliation[e]{Institut für Theoretische Physik, Universität Heidelberg, Philosophenweg 16, 69120 Heidelberg, Germany}

\emailAdd{alfio.bonanno@inaf.it}
\emailAdd{roberto.casadio@bo.infn.it}
\emailAdd{a.platania@thphys.uni-heidelberg.de}

\arxivnumber{1910.11393}

\abstract{A new class of relativistic stellar structure equations {which include the effects of an energy-dependent Newton coupling} is presented. Significant modifications in the mass-radius relation for neutron stars are possible only if the running of the Newton coupling due to quantum gravity occurs at low energies. A new Buchdahl limit is derived and its physical implications are discussed. 
In particular, sub-Planckian self-gravitating objects with arbitrarily small radii are possible.}

\begin{document}
\maketitle
\flushbottom

\section{Introduction} \label{intro} 

One of the most important issues in modern astrophysics is to understand the fate of stars undergoing a gravitational collapse.
It is believed that when the energy density of the collapsing star reaches Planckian scales, quantum effects will halt the collapse.
The details of this mechanism depend on the theory of quantum gravity at hand and, in recent years,
various scenarios have been put forward (see Ref.~\cite{Malafarina:2017csn} for a review). 

Over the past decades, Asymptotically Safe Gravity has emerged as a promising approach to obtain a consistent and predictive
theory of quantum gravity.
The key ingredient is the existence of a non-Gaussian fixed-point, acting as an ultraviolet (UV) attractor for the gravitational
renormalization group flow.
The possibility of quantizing gravity along these lines was first suggested by Weinberg in his seminal study in $d=2+\epsilon$
dimensions~\cite{1979W}, but at that time this fixed point was not accessible in $d=4$ dimensions due to its non-perturbative
character. 
After the introduction of the functional renormalization group equation for the effective average action~\cite{Wetterich:1992yh,Reuter:1996cp}, a systematic investigation of the phase diagram of gravity has become possible. 
The initial analyses based on the simple Einstein-Hilbert truncation~\cite{Souma:1999at} have been extensively generalized
by the inclusion of higher-derivatives operators~\cite{Lauscher:2002sq,Codello:2006in,Machado:2007ea,Benedetti:2009rx,
Falls:2016msz,Gies:2016con,Hamada:2017rvn,deBrito:2018jxt} and matter fields~\cite{Dona:2013qba,Dona:2014pla,
Meibohm:2015twa,Oda:2015sma,Dona:2015tnf,Eichhorn:2016esv,Christiansen:2017cxa,Eichhorn:2018nda,Eichhorn:2019yzm}.
The presence of a non-Gaussian fixed point has further been tested in unimodular-gravity settings~\cite{Eichhorn:2013xr,
Benedetti:2015zsw,Eichhorn:2015bna,deBrito:2019umw} and for the case of spacetimes carrying a foliation structure~\cite{Manrique:2011jc,
Biemans:2016rvp,Biemans:2017zca,Houthoff:2017oam,Platania:2017djo,Knorr:2018fdu}.
Beyond corroborating the existence of the non-Gaussian fixed point, all of these investigations support the conclusion that the
dimension of the UV critical manifold is finite~\cite{Falls:2017lst,Falls:2018ylp}. 

One of the key features of the Asymptotic Safety scenario for quantum gravity is the anti-screening of the gravitational interaction at short distances. In fact, the existence of a non-trivial fixed point for the dimensionless Newton coupling, $g(k)=G(k)\,k^2$, implies that the dimensionful Newton coupling vanishes at high energies, $G(k\to \infty)\to 0$. Implications of this anti-screening behaviour for the structure of black holes have appeared in Refs.~\cite{fayos11,Bonanno:2000ep,Bonanno:2006eu,Falls:2012nd,Koch:2013owa,Kofinas:2015sna,Bonanno:2016dyv,Bonanno:2017kta,Bonanno:2017zen,Torres:2017ygl,Pawlowski:2018swz,Adeifeoba:2018ydh,
Held:2019xde,Platania:2019kyx}, while consequences for the physics of the Early Universe have been studied in Refs.~\cite{irfp,br02,guberina03,Reuter:2005kb,br07,weinberg10,alfio12,saltas12,Copeland:2013vva,Kofinas:2016lcz,Bonanno:2015fga,Bonanno:2016rpx,
Bonanno:2017gji,Bonanno:2018gck,Platania:2019qvo}  (see Ref.~\cite{Bonanno:2017pkg} for a review).

The Asymptotic Safety mechanism can be understood in simple physical terms. Assuming that, in the large-distance limit, the leading quantum effects can be described by quantizing the linear fluctuations of the metric, one obtains a free field theory in a curved background spacetime whose elementary quanta, the gravitons, carry energy and momentum.
The vacuum of this theory is thereby populated by virtual graviton pairs, and a test body will become ``dressed'' by a cloud of virtual gravitons surrounding it. Whereas quantum fluctuations screen external charges in QED, they have an {\it antiscreening} effect on external
test masses in quantum gravity. 
As a consequence, the Newton constant becomes a density-dependent quantity which is smaller at shorter distances.
This behavior is similar to the running of the non-Abelian gauge coupling in Yang-Mills theories, for which the vacuum turns out to be ``paramagnetic" at large distances~\cite{Savvidy:1977as,Nink:2012vd}. 

As proposed by Markov and Mukhanov~\cite{marknc},  a way to consistently incorporate the antiscreening character of the gravitational interaction in Cosmology consists of introducing an energy-dependent Newton coupling as an effective multiplicative coupling between matter and geometry. In this work, we extend this formalism to study spherically symmetric field configurations. In particular we derive a modified 
Tolman-Oppenheimer-Volkoff (TOV) equation~\cite{Tolman:1939jz,Oppenheimer:1939ne,stephani}, and discuss its physical consequences. The specific scaling of the running gravitational coupling derived from the renormalization group analysis is then used at a later stage as an external input to work out explicit results, but different  scenarios for quantum gravity could also be considered (see, {e.g.},~Refs.~\cite{Sakharov:1967pk,Padmanabhan:2014jta,
Verlinde:2016toy,Dvali:2013eja,Giusti:2019wdx,Cadoni:2017evg,Cadoni:2018dnd,Casadio:2017twg,
Casadio:2019cux,Casadio:2018qeh,Carballo-Rubio:2017tlh}).

As expected, if quantum gravity sets in at Planckian scales, corrections to the mass-radius relation for macroscopic astrophysical sources are negligible. Clearly, these corrections would become phenomenologically more relevant in scenarios where the scale of quantum gravity is significantly lower than the Planck scale, but we do not pursue this line in the present work. The most significant  consequence of the energy dependence of the gravitational interaction is a modification of the Buchdahl bound \cite{Buchdahl:1959zz} for Planckian-size objects. Specifically, due to the antiscreening character of gravity at short distances, the Buchdahl limit deviates from the classical one at Planckian scales and matches the Schwarzschild limit at a critical radius  $R_\ast^\mathrm{crit}\sim0.37\, L_\mathrm{Pl}$, with $L_\mathrm{Pl}$ being the Planck length. Above this critical point, i.e., for $R\gtrsim R_\ast^\mathrm{crit}$, the formation of Planckian stars of compactness beyond the classical Buchdahl limit and positive central pressure is possible.
These Planckian-size stars are characterized by a radius extremely close to the Schwarzschild radius and, as a consequence, they would appear dark to distant observers. These objects thus match the idea of quasi-black-holes introduced in Refs.~\cite{Lemos:2007yh,Barcelo:2007yk} in the context of semi-classical gravity.
Below the critical point $R_\ast^\mathrm{crit}$, the quantum-improved Buchdahl limit undergoes a further deviation,
indicating a possible transition to a regime where the internal structure of stars is dominated by quantum-gravity fluctuations. 
In this regime, the existence of stars with positive isotropic internal pressure is more restricted, but the formation of horizonless astrophysical objects with arbitrarily small radii is still possible.

The plan of the paper is the following.
In Sect.~\ref{sec:grTOV}, we review the TOV equation in General Relativity.
The field equations in the Markov-Mukhanov formalism are obtained in Sect.~\ref{sec:fieldEq}, whereas the derivation of the improved
TOV equation is reported in Sect.~\ref{sec:impTOV}.
This equation is solved for a polytropic equation of state in Sect.~\ref{sec:results1}, and the corresponding modified mass-radius relation is discussed. The effects on the Buchdahl limit for uniform stars are studied in Sect.~\ref{sec:results2}.
Finally, in Sect.~\ref{sec:conc} we summarize our findings.

\section{TOV equation in Einstein gravity} 
\label{sec:grTOV}

In this section, we review the derivation of the classical TOV equation in General Relativity. This will serve to highlight the differences with respect to the case of an energy dependent coupling in the next section. 

A static and spherically symmetric astrophysical system, such as a star, is described by a metric of the form
\be \label{staticg}
\d s^2 =
c^2\, e^{\nu(r)}\,\d t^2 -e^{\lambda(r)}\,\d r^2 -r^2
\left(\d\theta^2+\sin^2\theta\,\d\phi^2\right)
\ .
\ee
The functions $\lambda(r)$ and $\nu(r)$ are determined by the internal structure of the star, specifically its composition
and the matter distribution in its interior.
A star can be modeled as a self-gravitating object made of a perfect fluid, with proper energy density $\epsilon=\epsilon(r)$
and hydrostatic pressure given by an equation of state (EoS) $p=p(\epsilon)$.
The energy-momentum tensor reads
\be \label{Tij}
T_{\mu\nu} =
\left[\epsilon+p(\epsilon)\right] c^{-2}\,u_\mu\,u_\nu - p(\epsilon)\,g_{\mu\nu}
\ .
\ee
Replacing Eq.~\eqref{staticg} and~\eqref{Tij} into the Einstein equations yields~\cite{stephani} 
\be
G^0_{\ 0}
&\equiv&
e^{-\lambda}
\left(\frac{\lambda'}{r}-\frac{1}{r^2}\right)
+
\frac{1}{r^2}
=
\kappa\,\epsilon
\label{G00}
\\
\nonumber
\\
G^1_{\ 1}
&\equiv&
-e^{-\lambda}
\left(\frac{\nu'}{r}+\frac{1}{r^2}\right)
+\frac{1}{r^2}
=
-\kappa\,p
\label{G11}
\\
\nonumber
\\
G^2_{\ 2} 
&\equiv&
-e^{-\lambda}
\left[\frac{\nu''}{2}+\frac{(\nu')^2}{4}-\frac{\nu'\,\lambda'}{4}
+\frac{\nu'-\lambda'}{2\,r}
\right] = -\kappa\,p
\label{G22}
\ ,
\ee
where  $f'=\partial_r f$ for any function $f$, $\kappa=8\pi G_0/c^4$, and the gravitational coupling is the constant $G_0$.
Since $\kappa$ is a constant, the conservation equation
$\nabla_\mu G^\mu_{\ \nu}=\kappa\,\nabla_\mu T^\mu_{\ \nu}=0$ reduce to the only non-trivial condition
\be
\label{cont}
\nabla_\mu T^\mu_{\ 1} \equiv -p'(r) - \frac{\nu'(r)}{2}\left(\epsilon+p\right)=0
\ .
\ee
Likewise, from Eq.~\eqref{G00}, one obtains 
\be
e^{-\lambda}=1-\frac{2\, G_0\, M(r)}{c^2\,r}
\ ,
\ee
where the mass function
\be
\label{massclass}
M(r) =\frac{4\,\pi}{c^2} \int_0^r \epsilon(x)\,x^2\,\d x
\ee
is such that the total mass of a star of radius $R_\ast$ is given by $M_\ast=M(R_\ast)$. This mass coincides with the Arnowitt-Deser-Misner~(ADM) mass~\cite{Arnowitt:1959ah} of the system. 

The classical TOV equation~\cite{Tolman:1939jz,Oppenheimer:1939ne,stephani} is finally obtained by combining
Eqs.~\eqref{G11} and~\eqref{cont}, 
\be
\label{grTOV}
p'(r)
=
-\left[p(r)+\epsilon(r)\right]
\frac{G_0}{c^2\,r^2}\left[M(r)+\frac{4\,\pi\, r^3}{c^2}\,p(r)\right]
\left[1-\frac{2\,G_0\,M(r)}{c^2\,r}\right]^{-1}
 \ ,
\ee
and reduces to the Newtonian equation of hydrostatic equilibrium in the non-relativistic limit, $c\to\infty$.
\par
The TOV equation~\eqref{grTOV} and the relation~\eqref{massclass} for the mass function are the equations
governing the hydrostatic equilibrium for a self-gravitating spherically symmetric object.
Once an EoS is specified, the TOV equation can be solved in order to determine the radial variation
of the internal pressure $p=p(r)$ in a star interior. 

\section{Field equations in the Markov-Mukhanov formalism}
\label{sec:fieldEq}
We next consider the case in which matter couples with gravity via an energy dependent coupling.
In order to derive the corresponding effective field equations, we follow the Markov-Mukhanov
formalism~\cite{marknc},
which was originally introduced in order to study a particular case in which the
energy-dependent effective Newton coupling was assumed to vanish at short distances.
Although the antiscreening character of gravity was introduced in Ref.~\cite{marknc} as an \textit{ad hoc\/}
assumption, it turns out to be realized if gravity is asymptotically safe~\cite{Niedermaier:2006wt}, 
{\rm i.e.}, if the gravitational RG flow attains an interacting fixed point at high energies.

The starting point is a fluid whose gravitational dynamics is governed by the action
\be
\label{effS}
S = \frac{1}{2\kappa} \int \d^4 x\,\sqrt{-g}
\left[R+2\,\chi(\epsilon)\,\mathcal{L} \right]
\ ,
\ee
where $\mathcal{L}=-\epsilon$ is the matter Lagrangian~\cite{Minazzoli:2012md} and $\epsilon$ is again the proper energy density of the fluid. The function~$\chi=\chi(\epsilon)$ is an effective multiplicative coupling encoding the way matter interacts with gravity. The metric variation of the matter part of the Lagrangian yields
\be
\frac{1}{\sqrt{-g}}\,\delta
\left(2\,\sqrt{-g}\,\chi\,\epsilon \right) =
2\left(\epsilon\,\frac{\partial\chi}{\partial\epsilon}+\chi\right)
\delta\epsilon -\chi\,\epsilon\,g_{\mu\nu}\,\delta g^{\mu\nu}
\ .
\ee
Here, the variation $\delta\epsilon$ is given by
\be \label{deps}
\delta\epsilon =
\frac{1}{2} \left[\epsilon+p(\epsilon) \right]
\left(g_{\mu\nu} - c^{-2} u_\mu\,u_\nu \right) \delta g^{\mu\nu}
\ ,
\ee
as recalled in Appendix~\ref{aV}. Upon varying the action~\eqref{effS} with respect to the metric, we obtain the modified Einstein equations
\be \label{efe}
R_{\mu\nu}-\frac{1}{2}\,R\,g_{\mu\nu} = \Lambda_{\mu\nu}
\ ,
\ee
where $\Lambda_{\mu\nu}$ is the effective energy-momentum tensor  
\be\label{emt}
\Lambda_{\mu\nu} \equiv
\left(\epsilon\,\frac{\partial\chi}{\partial\epsilon}+\chi\right)
T_{\mu\nu}
-\epsilon^2\,\frac{\partial\chi}{\partial\epsilon}\,g_{\mu\nu}
\ .
\ee
We see that the matter fields, whose energy-momentum tensor $T_{\mu\nu}$ is given by Eq.~\eqref{Tij},
couple to gravity via the effective Newton coupling
\be \label{stra}
G_\mathrm{eff}(\epsilon) \equiv
\frac{c^4}{8\,\pi}\,\frac{\partial(\epsilon\,\chi)}{\partial\epsilon}
\ .
\ee
In addition, the variation of the action~\eqref{effS}  generates an effective cosmological constant
\be
\Lambda_\mathrm{eff}(\epsilon)
=
-\epsilon^2\,\frac{\partial\chi}{\partial\epsilon}
\ .
\ee
One should further note that consistency with the Bianchi identity demands that $\nabla_\mu \Lambda^{\mu\nu}=0$
and, on considering the explicit expression~\eqref{Tij} of the matter energy-momentum tensor, it implies
\be \label{cc1}
\nabla_\mu
\Lambda^{\mu\nu}
=
\frac{\partial(\epsilon\,\chi)}{\partial\epsilon}\,
\nabla_\mu
T^{\mu\nu}
+
\frac{\partial^2(\epsilon\,\chi)}{\partial\epsilon^2}\,
(\epsilon+p)
\left(
c^{-2}u^\mu\,u^\nu-g^{\mu\nu}
\right)
(\partial_\mu\epsilon)=0
\ .
\ee
This constraint is satisfied if the usual conservation equation $\nabla_\mu T^{\mu\nu}=0$ holds,
along with (at least) one of the following conditions:
 \begin{enumerate} 
\item
\label{c1}
The quantity $(\epsilon\,\chi)$ is a linear function of the energy density:
\be
\frac{\partial^2(\epsilon\,\chi)}{\partial\epsilon^2}=0 \ .
\ee
In this case we must have $\chi = \chi_0+\frac{\chi_1}{\epsilon}$, 
which implies that $\chi\to \chi_0$ when $\epsilon\to \infty$.
Of course, a regular low energy limit ($\epsilon\to 0$) requires $\chi_1=0$, and one is left with
the standard general relativistic case upon identifying $\chi_0\equiv\kappa$ (see Eq.~\eqref{stra}).
\item
\label{c2}
The energy density and pressure satisfy the vacuum equation of state:
\be
\label{E-p}
\epsilon+p = 0
\ ,
\ee
meaning that the energy-momentum tensor is trivially conserved in de~Sitter spacetimes, independently
of the effective gravitational coupling $\chi$.
\item
\label{c3}
The gradient of the energy density is proportional to the fluid's four-velocity:
\be
\left(c^{-2}\,u^\mu\,u^\nu-g^{\mu\nu}\right) (\partial_\mu\epsilon) = 0
\ .
\ee
This case could be relevant for cosmology since it implies that the energy-momentum tensor
of a homogeneous fluid sourcing the Friedmann-Lemaitre-Robertson-Walker spacetime is generally
conserved for any choice of $\chi$.
 \end{enumerate} 
Conversely, if none of the above conditions holds, one cannot have $\nabla_\mu T^{\mu\nu}=0$, and
Eq.~\eqref{cc1} becomes a non-trivial constraint, which we can write as
\be
\nabla_\mu
\Lambda^{\mu}_{\ \nu}
=
\frac{\partial(\epsilon\,\chi)}{\partial\epsilon}\,
\nabla_\mu T^{\mu}_{\ \nu} + \frac{\partial^2(\epsilon\,\chi)}{\partial\epsilon^2}\,
(\epsilon+p) \left(c^{-2} u^\mu\,u_\nu-\delta^{\mu}_{\ \nu}
\right) (\partial_\mu\epsilon)
= 0 \ .
\label{RGcons}
\ee
In the following we will use the modified Einstein equations~\eqref{efe} and the conservation
equation~\eqref{RGcons} in order to derive the modified TOV equation. 
\section{Quantum-improved TOV equation}
\label{sec:impTOV}
We first notice that, for a static fluid of the kind considered in Sect.~\ref{sec:grTOV},
the four-velocity can be written as $u^\mu=(e^{\nu/2},0,0,0)$, and we can still assume a metric
of the form~\eqref{staticg}.
As a first step, we need to determine the functions $\lambda=\lambda(r)$ and $\nu=\nu(r)$ in
the line element~\eqref{staticg}.
To this end, we study the modified Einstein equations~\eqref{efe}. 
The first component of Eq.~\eqref{efe} reads
\be
G^0_{\ 0}
=
e^{-\lambda} \left(\frac{\lambda'}{r}-\frac{1}{r^2}\right) + \frac{1}{r^2}
=
\frac{8\,\pi\, G_\mathrm{eff}(\epsilon)}{c^4}\,\epsilon
-
\Lambda_\mathrm{eff}(\epsilon)
\equiv
\chi \epsilon
\ ,
\ee
and determines the $g_{11}$ component of the metric
\be
\label{lambdafunct}
e^{-\lambda}
=
1-\frac{2\, G_0\, M_{\rm q}(r)}{c^2\,r}
\ ,
\ee
where $M_{\rm q}=M_{\rm q}(r)$ is a modified mass function.
Specifically, introducing the effective proper energy density $\epsilon_{\rm q}\equiv \kappa^{-1}\,\chi\, \epsilon$, the quantum-corrected gravitational mass~$M_{\rm q}$ is written in the same form as in the classical case
\be
\label{RGmass}
M_{\rm q}(r)
=
\frac{4\,\pi}{c^2} \int_0^r \epsilon_{\rm q}(x) \,x^2 \,dx
\ ,
\ee
with the proper energy density $\epsilon$ replaced by its ``screened'' counterpart $\epsilon_{\rm q}$.
In terms of the effective energy density~$\epsilon_{\rm q}$ the scale-dependent Newton coupling reads
\be \label{screening}
G_\mathrm{eff}(\epsilon)
=
\left(\frac{d \epsilon_{\rm q}}{d \epsilon}\right)
G_0
\ee
and reduces to the classical Newton constant in the limit $\epsilon_{\rm q}\to \epsilon$, {i.e.}, 
for $\chi\to\kappa$.
Replacing Eq.~\eqref{lambdafunct} into the second component of the Einstein equations,
\be
G^1_{\ 1}
=
-e^{-\lambda} \left(\frac{\nu'}{r}+\frac{1}{r^2}\right) +\frac{1}{r^2}
=
-\left(\epsilon\,\frac{\partial\chi}{\partial\epsilon}+\chi\right)
p -\epsilon^2\,\frac{\partial\chi}{\partial\epsilon}
=
-\chi\,p - \epsilon\,\frac{\partial\chi}{\partial\epsilon}
(p+\epsilon)
\ ,
\ee
yields a differential equation for the function $\nu=\nu(r)$ which determines the temporal component of the metric,
\be
\label{eqnu}
\nu'(r)
=
\frac{2\,G_0}{c^2\,r^2}
\left\{M_{\rm q} +\frac{4\,\pi\, r^3}{c^2}\left[(p+\epsilon)
\left(\frac{d \epsilon_{\rm q}}{d \epsilon}\right) -\epsilon_{\rm q} \right]\right\}
\left(1-\frac{2\,G_0\,M_{\rm q}}{c^2\,r}\right)^{-1}
\ .
\ee
At this point, the modified TOV equation can be obtained straightforwardly from the conservation equation~\eqref{RGcons}.
The first component reads
\be
\nabla_\mu \Lambda^{\mu}_{\ 0}
=
\kappa\left(\frac{d \epsilon_{\rm q}}{d \epsilon}\right)
\nabla_\mu T^{\mu}_{\ 0}
=
0
\ ,
\label{RGcons0}
\ee
and is identically satisfied (like in the standard case), whereas 
\be
\nabla_\mu
\Lambda^{\mu}_{\ 1}
=
-\kappa
\left(\frac{d \epsilon_{\rm q}}{d \epsilon}\right)
\left[p'+\frac{\nu'}{2}\left(p+\epsilon\right)\right]
-\kappa\left(\frac{d^2\epsilon_{\rm q}}{d \epsilon^2}\right)
\left(p+\epsilon\right)
\epsilon'
=
0
\ ,
\label{RGcons1}
\ee
differs from Eq.~\eqref{cont}, because of the dependence of the effective gravitational coupling
on the proper energy density,~$\chi=\chi(\epsilon)$.
From the above equation, we finally obtain 
\be
\label{rgTOV}
p'(r)
=
-(p+\epsilon)
\left\{
\frac{\nu'(r)}{2}+\epsilon'(r)\frac{d}{d\epsilon} \left[\mathrm{log}\left(\frac{d \epsilon_{\rm q}}{d\epsilon}\right)\right]
\right\}
\ .
\ee
This relation, combined with Eq.~\eqref{eqnu}, yields the modified TOV equation in terms of the effective energy density
$\epsilon_{\rm q}$ and mass $M_{\rm q}$.
As a check, we notice that Eqs.~\eqref{eqnu} and \eqref{rgTOV} give Eq.~\eqref{grTOV} when $\epsilon_{\rm q}\to\epsilon$,
or $\chi=\kappa=\text{constant}$.

In the next sections, we will discuss the modifications to the mass-radius relation, maximal mass, and Buchdahl limit induced by the modified TOV equation~\eqref{rgTOV}.

\section{Mass-radius relation for polytropic {quantum-improved} neutron stars} \label{sec:results1}

In this section we study the structure of astrophysical compact objects (e.g., neutron stars), arising from the modified TOV equation derived in the previous section.
Classically, once an EoS is specified, the mass-radius relation arising from the TOV equation leads to an upper bound for the mass of neutron stars \cite{Ozel:2012ax,Chamel:2013efa,Margalit:2017dij,Cromartie:2019kug}.  
In what follows we will discuss how the antiscreening character of gravity predicted in the AS scenario could modify the mass-radius relation and the corresponding maximal mass.

In order to solve the new TOV equation, we need to know how quantum gravitational effects modify the mass of a neutron star via the effective energy density $\epsilon_{\rm q}$. Solving the beta functions for the Newton coupling in the Einstein-Hilbert truncation~\cite{Bonanno:1998ye} gives
\be
\label{newton}
G(k) = \frac{G_0}{1+g_\ast^{-1}\, \xi^2\, (k/k_\mathrm{Pl})^2}
\ ,
\ee 
where $k_\mathrm{Pl}^2=\hbar \,c^3/G_0\equiv (M_\mathrm{Pl}\,c)^2$, $g_\ast\neq0$ is the fixed-point value attained by
the dimensionless running Newton coupling $g(k)\equiv G(k)\,k^2$ at high energies and $\xi$ is a positive constant setting
the scale at which quantum gravitational effects become important.
Specifically, the scale of quantum gravity is set by the momentum scale
\be
k_{\rm tr}=\xi^{-1}k_\mathrm{Pl} \ ,
\ee
and it gives the Planck scale for $\xi\sim\mathcal{O}(1)$.
Note that the Newton coupling~\eqref{newton} vanishes when $k\to\infty$, and in particular $G(k)\leq G_0$ at all scales.
This is reminiscent of the antiscreening behaviour of gravity described in the introduction.
\par
With the expression~\eqref{newton} for the running Newton coupling and the cutoff function
$k^2=\frac{\hbar^2 \,G_0}{c^4}\, \epsilon$~\cite{Platania:2019kyx}, integrating Eq.~\eqref{screening} yields 
\be
\label{rescreeneps}
\epsilon_{\rm q}(\epsilon)
=
\frac{{\log}\left(1+g_\ast^{-1}\,\xi^2 \, \epsilon/\epsilon_\mathrm{Pl}\right)}{g_\ast^{-1}\, \xi^2 / \epsilon_\mathrm{Pl}}
\ ,
\ee
where $\epsilon_\mathrm{Pl}\equiv \frac{c^7}{\hbar \, G_0^2}$ is the Planck density. The latter enters the expression of the effective energy density $\epsilon_{\rm q}$ as
\be
\label{suppr}
\epsilon_\mathrm{Pl}^{-1}
\simeq
1.176\cdot 10^{-76}\, \frac{r_g^2}{M_\odot\, c^2}
\ ,
\ee
where $r_g\equiv \frac{G_0\,M_\odot}{c^2}\simeq 1.48 \,\mathrm{km}$ is half of the Schwarzschild radius of the sun.
The effective energy density $\epsilon_{\rm q}$ reduces to the classical proper energy density
$\epsilon$ in the limit~$\xi\to0$, as expected. 
The expression~\eqref{rescreeneps} for the effective proper energy density affects both the continuity equation~\eqref{RGmass} and the TOV equation~\eqref{rgTOV} and is the only
input needed to embed the scale-dependence of the running Newton coupling into the dynamics of the system. 

We preliminary notice that, due to the antiscreening of gravity at high energies, the mass $M_{\rm q}$ will be smaller than its classical counterpart. In fact, the quantum-improved mass~\eqref{RGmass} satisfies the following inequality
\be
M_{\rm q}(r)
=
\frac{4\,\pi}{c^2} \int_0^r \epsilon_{\rm q}(x) \, x^2\, dx
=
\frac{4\,\pi}{c^2}\int_0^r
\left[\int_0^{\epsilon(x)}\frac{G(\epsilon')}{G_0}\,d\epsilon'\right]
x^2\,dx 
\leq \frac{4\,\pi}{c^2} \int_0^r \epsilon(x) \, x^2\, dx
= M(r) \ .
\ee
Therefore, neutron stars formed in the presence of an energy-dependent Newton coupling, which decreases at short distances, would be lighter than classical neutron stars.

The mass-radius relation for a neutron star can now be obtained by replacing the expression
for $\epsilon_{\rm q}$ in the modified TOV equation~\eqref{rgTOV}, and by specifying an EoS.
In the following, we will only consider a polytropic fluid
\be
\label{eospoly}
p(r)=\gamma \,\epsilon(r)^{1+\alpha}
\ ,
\ee
with polytropic index $\alpha=1$ and $\gamma\sim 4\cdot10^{-4}\,\mathrm{fm}^3/\mathrm{MeV}$~\cite{Silbar:2003wm}.
This EoS is supposed to mimic the effects of strong nucleon-nucleon interactions in a neutron star interior~\cite{Silbar:2003wm}.
Finally, the integration of the TOV equation, together with the conditions~$\epsilon(R_\ast)=0$ and
$\epsilon(0)\equiv \epsilon_0$, allow us to compute the modifications of the mass-radius relation~$M_\ast(R_\ast)$ induced by quantum-gravity effects in a neutron-star interior.
In particular, for a given central density $\epsilon_0$, the radius is obtained by the condition $p(R_\ast)=\epsilon(R_\ast)=0$,
while the quantum-corrected gravitational mass is defined by~$M_\ast\equiv M_{\rm q}(R_\ast)$.
\begin{figure}[t!]
    \centering
    \includegraphics[width=0.58\textwidth]{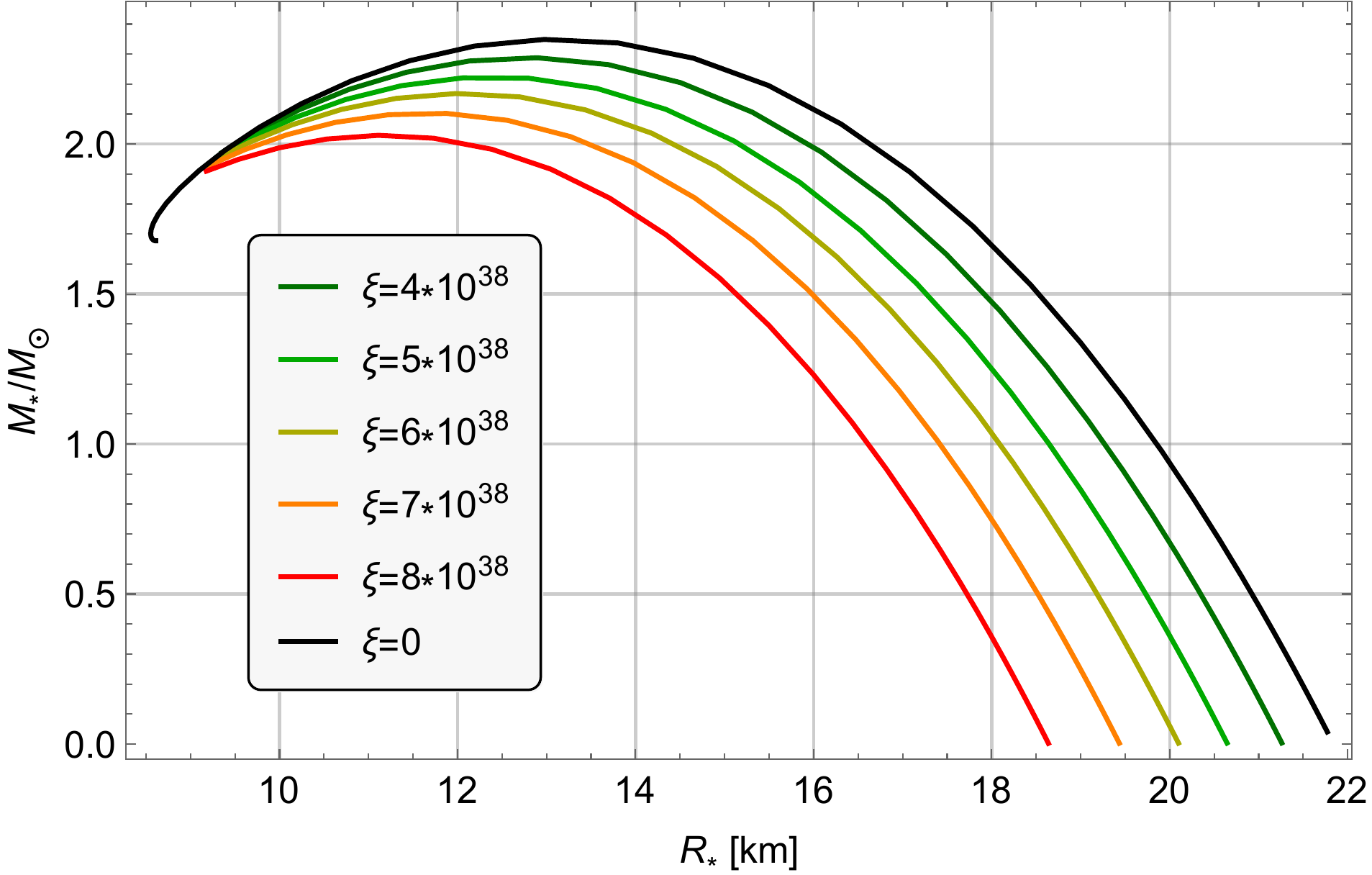}
    \caption{Mass-radius relation for various values of the parameter~$\xi$ and central density~$10^{-6}\le \epsilon_0\le 0.01\,M_\odot c^2 / r_g^3$. {The classical case corresponds to~$\xi=0$ (black line).
    If quantum-gravity effects take place at the Planck scale, {i.e.}~$\xi\sim\mathcal{O}(1)$, modifications to the mass-radius relation are strongly suppressed. In order to get a sizable modification at macroscopic scales, one needs~$\xi\gtrsim 10^{38}$, and the antiscreening character of gravity then leads to the existence of lighter and smaller neutron stars.}}
    \label{fig:1}
\end{figure}
\par
\begin{figure}[t!]
    \centering
    \includegraphics[width=0.47\textwidth]{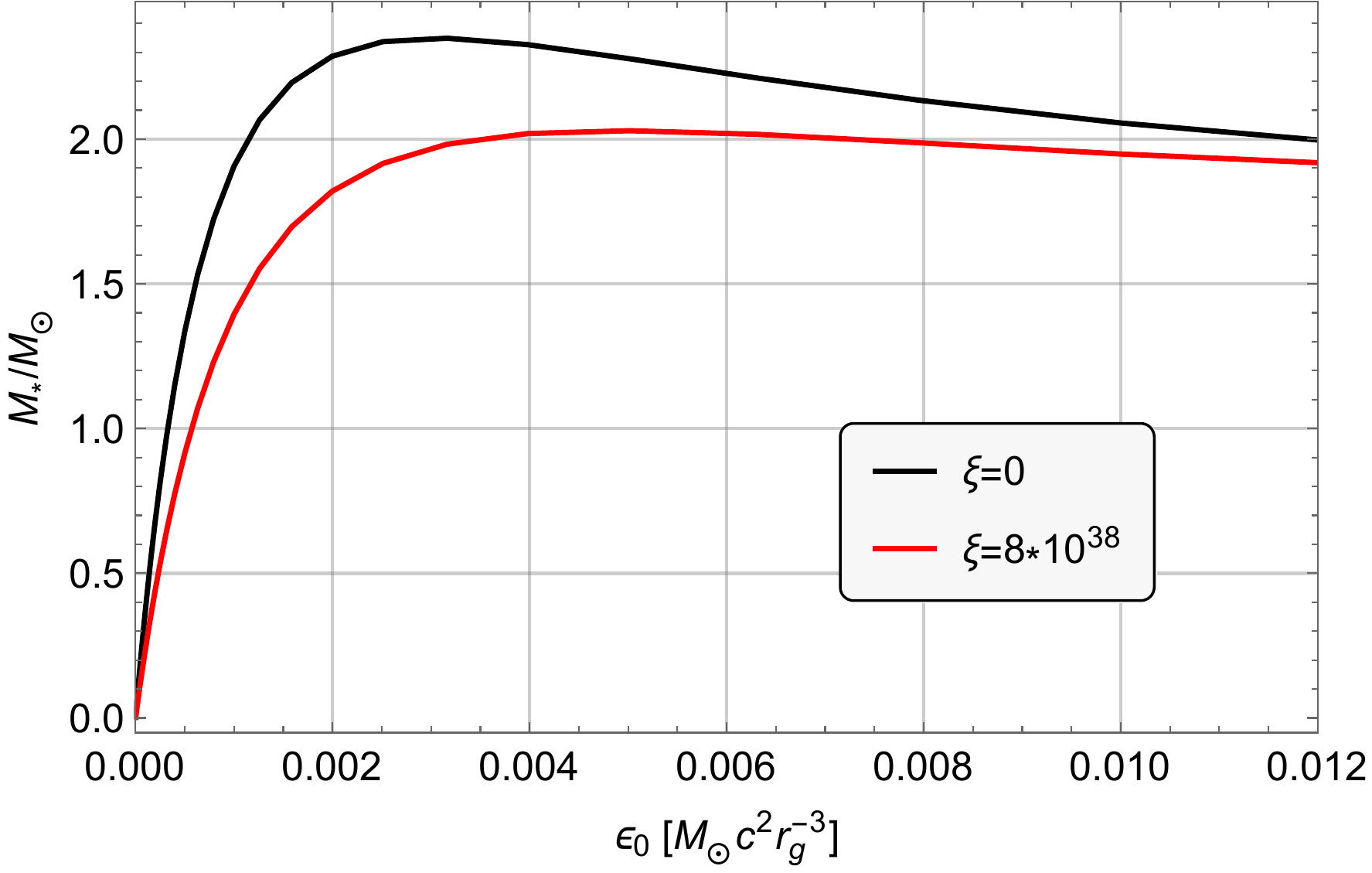} $\quad$
    \includegraphics[width=0.47\textwidth]{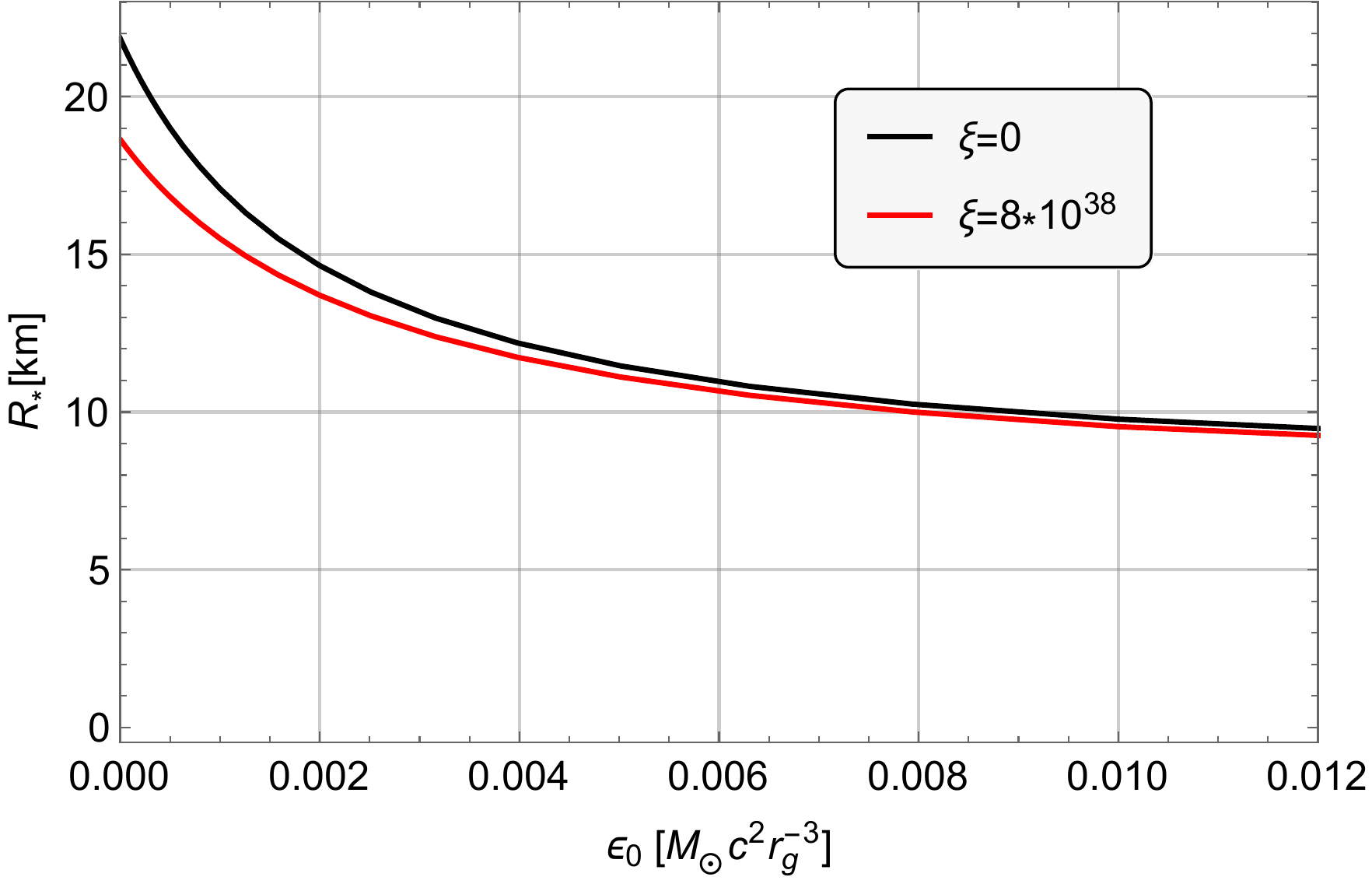}
    \caption{Dependence of mass $M_*$ (left panel) and radius $R_*$ (right panel) on the central density
    $\epsilon_0$.
    The black line ($\xi=0$) reproduces the classical case, whereas the red line is obtained in the quantum-modified
    case for $\xi=8\cdot 10^{38}$.}
    \label{fig:2}
\end{figure}

The mass-radius relation clearly depends on the scale at which the gravitational antiscreening becomes important.
As already mentioned, this scale is parametrised by the positive constant $\xi$. The modified mass-radius relation is shown in Fig.~\ref{fig:1} for different values of $\xi$. 
Unsurprisingly, a sizable effect on the mass and radius of a  star requires to raise the scale of quantum gravity to
macroscopic scales, by setting $\xi^2\gtrsim10^{76}$.
This enormous value for $\xi^2$ is due to the suppression factor $\epsilon_\mathrm{Pl}^{-1}$, Eq.~\eqref{suppr}, which enters the expression~\eqref{rescreeneps} for the effective energy density $\epsilon_{\rm q}$.
As already noticed, the antiscreening effects induced by the running gravitational coupling~\eqref{newton} decrease
the value of the effective mass $M_{\rm q}(R_\ast)$, thus favouring the formation of lighter and smaller neutron stars.
This is explicitly shown in Fig.~\ref{fig:2}, where the classical (black line, $\xi=0$) and quantum-modified
(red line, $\xi=8\cdot10^{38}$) mass and radius are shown as functions of the central density $\epsilon_0$.
%
%
%\begin{figure}[t!]
   % \centering
    %\includegraphics[width=0.47\textwidth]{DeltaM.pdf}$\quad$
    %\includegraphics[width=0.47\textwidth]{DeltaR.pdf}
    %\caption{Deviation of mass (left panel) and radius (right panel) of an \DIFdelbeginFL \DIFdelFL{RG-improved }\DIFdelendFL \DIFaddbeginFL \DIFaddFL{quantum-modified }\DIFaddendFL neutron star
    %($\xi=5\cdot 10^{77}$) from the corresponding classical quantities. Specifically, the plots show the dependence of the differences $\Delta M_\ast=M_\ast-M_c$ and $\Delta R_\ast=R_\ast-R_c$ on the central density $\epsilon_0${\color{magenta}, $(M_c,R_c)$ being the mass and radius obtained from the classical theory}. }
    %\label{fig:3}
%\end{figure}
%
%
%
\section{Quantum-corrected Buchdahl limit}
\label{sec:results2}
In this section we study the TOV equation for an incompressible fluid characterized by a constant energy density~$\epsilon=\epsilon_0$.
Although the energy density in a realistic star interior certainly depends on the radial coordinate $r$, assuming~$\epsilon=\epsilon_0$ has the advantage of not requiring an EoS as an input to solve the TOV equation and allows to derive an important theoretical bound, known as the Buchdahl limit, for stars with isotropic pressure~\cite{Buchdahl:1959zz}.
In General Relativity, the latter is parametrized by a curve in the $(M_\ast-R_\ast)$-plane corresponding to spherically symmetric self-gravitating objects with diverging central pressure.
In what follows we will discuss how this classical limit is modified by antiscreening effects of gravity.

\begin{figure}[t!]
     \centering
     \includegraphics[width=0.47\textwidth]{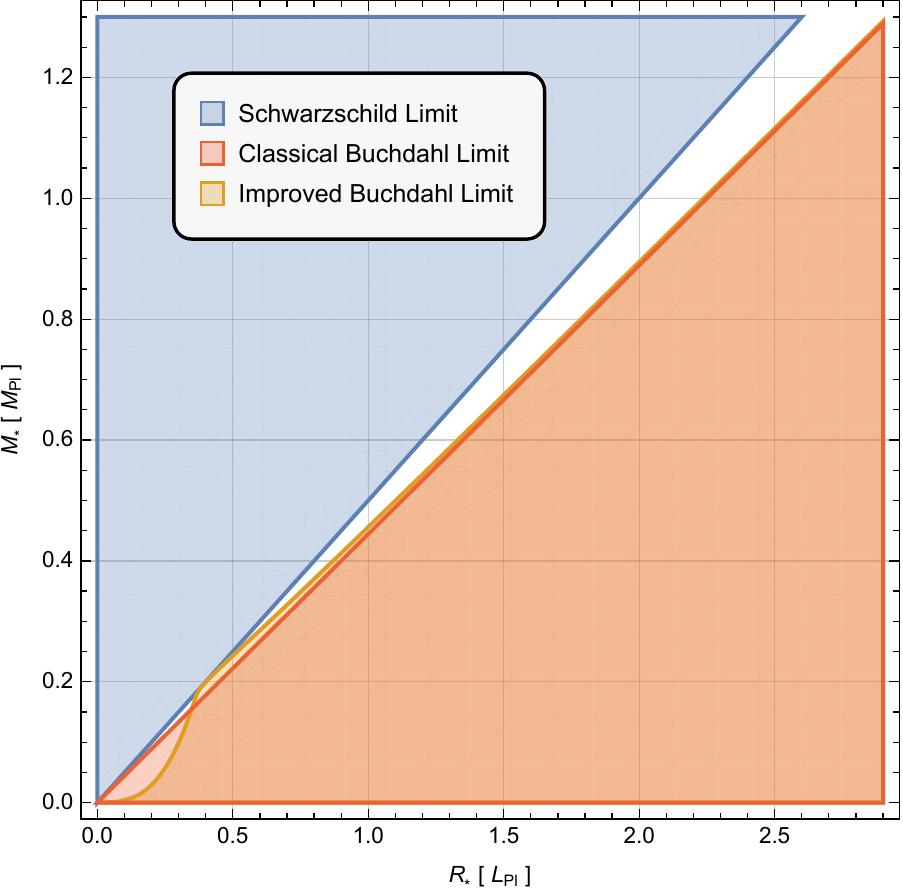}
    $\quad$\includegraphics[width=0.47\textwidth]{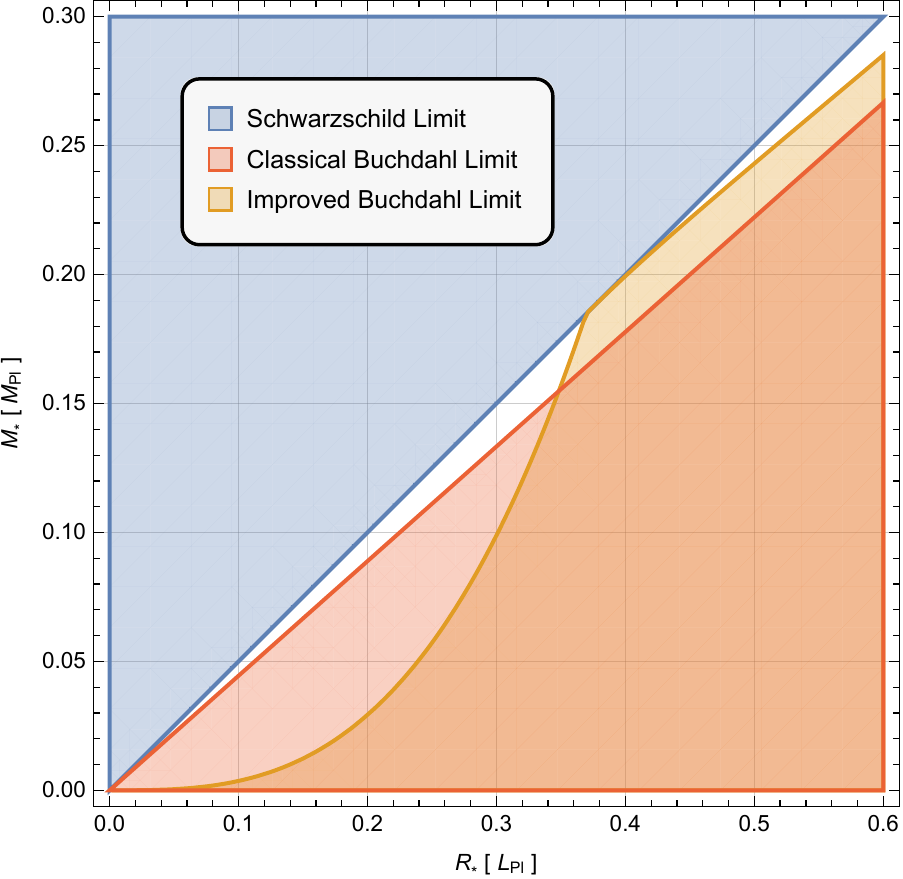}
    \caption{Graphical representation of the Buchdahl limit in the classical (orange region) and quantum-improved case (yellow region), for $\xi=g_\ast=1$, in Planck units. Astrophysical objects with radius $R_\ast<2\,M_\ast G_0/c^2$ fall into the blue region, delimited by the classical Schwarzschild limit. The white region in between is a transient region and accommodates astrophysical objects characterized by a negative central pressure. {For large masses and radii the classical and modified Buchdahl limits coincide, whereas modifications due to the running of the gravitational coupling occur at Planckian scales (see left panel). In particular, there exists a critical radius~$R_\ast\simeq0.37 \,L_{\rm Pl}$, where the improved Buchdahl-limit curve meets the classical Schwarzschild limit,~$R_\ast=2\,M_\ast\, G_0/c^2$. Below this critical point, i.e., for $R\lesssim R_\ast^\mathrm{crit}$, the internal structure of stars changes so that the Buchdahl limit is met when~$M_\ast \sim (R_\ast/L_\mathrm{Pl})^3 M_\mathrm{Pl}$, according to Eq.~\eqref{planckbuch}. A magnified view of the sub-Planckian region including the critical point is shown in the right panel.}
} \label{fig:4}
\end{figure}

We start by reviewing the derivation of the Buchdahl limit in the classical case. The mass of the star is obtained by integrating Eq.~\eqref{massclass} for a constant energy density
$\epsilon_0=\rho_0 \,c^2$,
\be
\label{massint}
M(r)
=
\frac{4\,\pi\, r^3}{3}\,\rho_0
=
\frac{M_*\,r^3}{R_*^3}
\ ,
\ee
where we used 
\be
\rho_0=\frac{3\, M_\ast}{4\,\pi\, R_\ast^3}
\label{rho0gr}
\ee
for the total mass density.
The isotropic hydrostatic pressure $p=p(r)$ is then derived by replacing Eq.~\eqref{massint} into the classical TOV
equation~\eqref{grTOV}, and solving with the boundary condition $p(R_\ast)=0$.
The central pressure reads 
\be
p(0)
=
-\frac{3\,c^2\,M_*\left[
3\,G_0\, M_\ast
+c^2 \,R_\ast\, \left(\sqrt{1-\frac{2\,G_0\, M_\ast}{R_\ast\,c^2}}-1\right)\right]}
{4\,\pi\, R_\ast^3\,(4\,c^2\,R_\ast-9\,G_0\,M_\ast)}
\ .
\ee
The condition $p^{-1}(r=0)>0$, ensuring a finite and positive central
pressure~\footnote{Note that the positive-pressure condition is not sufficient to determine the stability of a star.
In particular, in spite of their negative central pressure, ultra-compact Schwarzschild stars beyond the Buchdahl limit could be dynamically stable against radial perturbations~\cite{Camilo:2018goy}.},
thus imposes the Buchdahl bound
\be
\label{classbuchdahl}
M_\ast
<
\frac{4\,c^2}{9\,G_0}\,R_\ast
\ .
\ee
The allowed region in the $(M_\ast-R_\ast)$-plane is shown in the left panel of Fig.~\ref{fig:4} (orange region),
together with the region $R_\ast>2\,M_\ast\, G_0/c^2$ corresponding to astrophysical objects whose radius
is inside the Schwarzschild radius (blue region).
\par
Modifications induced by the energy dependence of the gravitational coupling can now be obtained by following the same steps.
We first need to solve Eq.~\eqref{RGmass} for a constant energy density $\epsilon_0=\rho_0\,c^2$.
The corresponding mass function is given by
\be
\label{genmass}
M_{\rm q}(r)
=
\frac{4\,\pi\, r^3}{3}\left[
\frac{{\log}\left(1+\xi^2 \, g_\ast^{-1} \epsilon_\mathrm{Pl}^{-1}\,\rho_0\, c^2\right)}
{\xi^2 \, g_\ast^{-1}\, \epsilon_\mathrm{Pl}^{-1}\,c^2}
\right]
\ ,
\ee
so that a star of radius $R_\ast$ and mass $M_\ast=M_{\rm q}(R_\ast)$ has an effective mass density
\be
\rho_0
=
\left[\exp\left(\frac{\xi^2 c^2}{g_\ast\,\epsilon_\mathrm{Pl}}\,\frac{3\,M_\ast}{4\,\pi\, R_\ast^3 }\right)-1 \right]
\frac{g_\ast\,\epsilon_\mathrm{Pl}}{\xi^2 c^2}
\ .
\label{rho0imp}
\ee
{Note that, for objects of small compactness, $G_0\,M_*\ll c^2\,R_*$, the above expressions just reproduce their general relativistic analogues, Eq.~\eqref{massint} and~\eqref{rho0gr}.
Employing Eqs.~\eqref{rho0imp} and~\eqref{genmass}, the new TOV equation~\eqref{rgTOV} can be integrated to give an explicit expression for the pressure of the star.} In the center, it is given by
\be
p(0)
=
\frac{\epsilon_\mathrm{Pl}}{\xi^2 g_\ast^{-1}}
\,\frac{\mathcal{N}(R_\ast,M_\ast)}{\mathcal{D}(R_\ast,M_\ast)}
\ ,
\ee
with
\begin{align}
\mathcal{N}
=&
\left(e^{\frac{\xi^2 c^2}{g_\ast\epsilon_\mathrm{Pl}}\,\frac{3\,M_\ast}{4\,\pi \,R_\ast^3 }}-1\right)
\left[\xi^2 g_\ast^{-1}M_\ast \, e^{\frac{\xi^2 c^2}{g_\ast\,\epsilon_\mathrm{Pl}}\frac{3\,M_\ast}{4\,\pi\, R_\ast^3 }}
-2\,\pi\, R_\ast^3\,\frac{\epsilon_\mathrm{Pl}}{c^2}
\left(e^{\frac{\xi^2 c^2}{g_\ast\,\epsilon_\mathrm{Pl}}\,\frac{3\,M_\ast}{4\,\pi\, R_\ast^3 }}-1\right)
\right] \nonumber \\ 
& \times \left(\sqrt{1-\frac{2\,G_0\, M_\ast}{R_\ast\,c^2}}-1\right)
\end{align}
and
\be
\mathcal{D}
=
\xi^2 g_\ast^{-1}\,M_\ast \, e^{\frac{\xi^2 c^2}{g_\ast\,\epsilon_\mathrm{Pl}}\,\frac{3\,M_\ast}{4\,\pi\, R_\ast^3 }}
+2\,\pi\, R_\ast^3\,\frac{\epsilon_\mathrm{Pl}}{c^2}
\left(e^{\frac{\xi^2 c^2}{g_\ast\,\epsilon_\mathrm{Pl}}\,\frac{3\,M_\ast}{4\,\pi\, R_\ast^3 }}-1\right)
\left(\sqrt{1-\frac{2\,G_0\, M_\ast}{R_\ast\,c^2}}-1\right).
\ee
The condition $p^{-1}(r=0)>0$ defines the quantum-improved Buchdahl bound (yellow region in Fig.~\ref{fig:4}).
For large masses and radii, the limiting curve $p^{-1}(r=0)=0$ coincides with the classical Buchdahl limit~\eqref{classbuchdahl}.
Zooming into the Planckian region, where ultra-compact stars of Planckian masses are located, this limiting curve starts deviating from the classical one (see left panel of Fig.~\ref{fig:4}), and gets closer to the Schwarzschild limit.
Our modified TOV equation thus allows for the existence of ultra-compact horizonless objects of positive central pressure beyond the classical Buchdahl limit. The radius of these Planckian stars is very close to the Schwarzschild radius. As a consequence, similarly to the ``quasi-black-holes'' introduced in Refs.~\cite{Lemos:2007yh,Barcelo:2007yk}, these ultra-compact objects would appear dark to distant observers. 

Interestingly, there exists a critical point $(R_\ast^\mathrm{crit},M_\ast^\mathrm{crit})$ where the modified Buchdahl bound matches the Schwartzschild limit.
The value of the critical radius~$R_\ast^\mathrm{crit}$ can be derived analytically from the condition~$\mathcal{D}(R_\ast,R_\ast c^2 /2 G_0)=0$ and reads
\be
R_\ast^{\rm crit}
=
L_\mathrm{Pl} \,\sqrt{\frac{\xi^2 \, g_\ast^{-1}}
{4\,\pi\left[1+\frac{2}{3}\mathcal{W}\left(-\frac{3}{2}\,e^{-3/2}\right)\right]}}
\simeq
0.37 \,\sqrt{\xi^2 \,g_\ast^{-1}} \,L_\mathrm{Pl}
\ ,
\ee
where $L_\mathrm{Pl}$ is the Planck length and $\mathcal{W}$ is the Lambert W-function.
The existence of a critical value is due to the presence of the non-trivial fixed point $g_\ast$, reminiscent of the non-perturbative renormalizability of gravity. Naturally, in the classical limit $g_\ast\to\infty$ (or, equivalently, $\xi\to0$) and the critical radius $R_\ast^{\rm crit}$ vanishes. 

Beyond the critical point, the Buchdahl limit undergoes a major deviation from the classical case, indicating a possible transition to a new phase, dominated by quantum-gravitational fluctuations.
In fact, crossing the critical point, the function $\mathcal{D}(R_\ast,M_\ast)$ becomes negative and the Buchdahl condition reduces to the inequality $\mathcal{N}(R_\ast,M_\ast)<0$.
The latter can be solved exactly and gives a quantum-improved Buchdahl limit for sub-Planckian stars,
\be
\label{newbuch}
M_\ast
<
\frac{2\,\pi\, c^5 \left[1+\frac{2}{3}\,\mathcal{W}\left(-\frac{3}{2}\,e^{-3/2}\right)\right]}
{\hbar\, \xi^2 g_\ast^{-1}\, G_0^2}\,R_\ast^3
\ .
\ee
The above equation is equivalent to the condition
\be
\label{planckbuch}
R_\ast\gtrsim\,0.27\,\xi^2 g_\ast^{-1}\left(\frac{M_\ast}{M_\mathrm{Pl}}\right)^{\frac{1}{3}}
L_\mathrm{Pl}  
\ee
and, interestingly, the critical line $R_\ast\sim (M_\ast/M_\mathrm{Pl})^{1/3}\,L_\mathrm{Pl}$ resembles the scaling relation characterizing ``Planck stars'' of the kind introduced in Ref.~\cite{Rovelli:2014cta}. 
As is clear from Eq.~\eqref{newbuch}, below the critical point the quantum-corrected Buchdahl limit deviates substantially from the classical one. This transition occurs in the sub-Planckian region, as shown in the right panel of Fig.~\ref{fig:4}. 
At variance with the case $R_\ast^\mathrm{crit}<R_\ast\lesssim L_\mathrm{Pl}$, below the critical point the formation of stars of positive central pressure is more restricted. However, the new bound~\eqref{planckbuch} still allows for the existence of astrophysical objects with arbitrarily small radii.

\section{Conclusions} \label{sec:conc}

The approach discussed in this paper allows the study of stellar structure equilibrium configurations in theories with energy-dependent gravitational interaction. This dependence could arise, e.g., from the running of the gravitational couplings dictated by the renormalization group equations.

Unless the scale of quantum gravity is much below the Planck scale, quantum-gravity effects cannot affect the mass-radius relation of standard neutron stars. However, our formalism is useful to describe possible outcomes of the gravitational collapse when the internal density reaches Planckian values. In particular the new, quantum-improved, stellar structure equations could be used to discuss possible non-singular stellar-like remnants. 
Clearly, in order to address the latter point, it is essential to study the stability of these new configurations. Classical secular stability can be discussed in terms of the critical points of the mass $M_\ast(\rho_0)$, where $\rho_0$ is the central mass density. 
In our case, unless the running of Newton coupling is negligible, this criterion cannot be used. It is thereby essential to study the spectrum of the stability equations for the radial modes induced by the effective energy-momentum tensor~\eqref{emt}. We defer this point 
to a following paper. 

Assuming that the running of the Newton coupling is given by Eq.~\eqref{newton} and a polytropic equation of state (our arguments can be easily generalized to the case of piecewise equation of states), the most important result of our investigation is the possible existence of ultra-compact objects of Planckian size, as is displayed in Fig.~\ref{fig:4}. These Planckian stars would be completely dark for any astrophysical purpose, yet they would have no horizon. 
They could have been produced by direct collapse of primordial perturbations during the inflationary era and 
subsequently merged to produce primordial black holes~\cite{Carr:1974nx,Carr:1975qj,Carr:2016drx,Casadio:2018vae}.
It would thus be interesting to study the phenomenological consequences of our findings.

\acknowledgments{A.B.~and R.C.~are partially supported by the INFN grant FLAG.
The work of R.C.~has also been carried out in the framework of activities of the
National Group of Mathematical Physics (GNFM, INdAM) and COST action {\em Cantata\/}.
The research of A.P.~is supported by the Alexander von Humboldt foundation.}

\appendix
\section{Varying the matter Lagrangian} \label{aV}
\setcounter{equation}{0}
In this Appendix, we derive the variation~\eqref{deps} for the proper energy density following
closely Ref.~\cite{Harko:2010zi}. Our convention for the metric signature is $(+---)$.

Let us consider a generic fluid of baryonic density $\rho$.
The baryonic density is a density function satisfying the continuity equation for vanishing pressure
\be
\nabla_\mu
\left(\rho\,u^\mu
\right)
=
\frac{1}{\sqrt{-g}}\,
\partial_\mu
\left(\sqrt{-g}\,\rho\,u^\mu
\right)
=
0 \ ,
\label{cx}
\ee
where $u^\mu$ is the fluid's 4-velocity in a general reference frame $x^\mu$.
Specifically, $u^\mu=\frac{\d x^\mu}{\d\tau_x}$, with the proper time implicitly defined by
\be \label{propertime}
c^2 \d\tau_x^2
=
g_{\mu\nu}(x^\sigma)\,\d x^\mu\,\d x^\nu
\ .
\ee
The relation~\eqref{cx} depends on the metric, both explicitly via the determinant $g$
and implicitly in the definition~\eqref{propertime} of proper time $\tau_x$.
However, $\rho$ is a scalar under change of coordinates, so that the same continuity equation must also hold in the frame $X^\alpha=X^\alpha(x^\mu)$ comoving with the fluid. In this frame, the fluid's proper time along~a trajectory $X^\alpha=X^\alpha(\tau_X)$ is given by
\be
c^2\d\tau_X^2
=
G_{00}(X^\alpha)\,\d X^0\,\d X^0
=
g_{\mu\nu}(x^\sigma)\,\frac{\partial x^\mu}{\partial X^0}\,\frac{\partial x^\nu}{\partial X^0}\,
\d X^0\,\d X^0
\ ,
\ee
and the 4-velocity is
\be
U^\alpha
=
\frac{\d X^\alpha}{\d\tau_X}
=
\frac{\delta^\alpha_{0}}{c^{-1}\left[
g_{\mu\nu}\,\frac{\partial x^\mu}{\partial X^0}\,\frac{\partial x^\nu}{\partial X^0}
\right]^{1/2}}
\ .
\ee
The continuity equation in the comoving frame reads
\be
\frac{1}{\sqrt{-G}}\,
\partial_0
\left(\sqrt{-G}\,\rho\,U^0
\right)
=
0
\ ,
\label{cX}
\ee
and can be integrated to yield
\be
\sqrt{-g}\,\left|\frac{\partial x}{\partial X}\right|
\,\rho
=
\sqrt{-G}\,\rho
=
K\,\left[
g_{\mu\nu}\,\frac{\partial x^\mu}{\partial X^0}\,\frac{\partial x^\nu}{\partial X^0}
\right]^{1/2}
\ ,
\label{K}
\ee
where $K$ is an integration constant and $\left|\frac{\partial x}{\partial X}\right|$
the modulus of the Jacobian of the transformation $x^\mu=x^\mu(X^\alpha)$, which does
not depend on the metric.
It follows that 
\be
\delta\left(\sqrt{-g}\,\left|\frac{\partial x}{\partial X}\right|
\,\rho\right)
=
\left|\frac{\partial x}{\partial X}\right|
\delta\left(\sqrt{-g}\,\,\rho\right)
=
\frac{K}{2}\left[
g_{\mu\nu}\,\frac{\partial x^\mu}{\partial X^0}\,\frac{\partial x^\nu}{\partial X^0}
\right]^{-1/2}
\frac{\partial x^\sigma}{\partial X^0}\,\frac{\partial x^\lambda}{\partial X^0}
\,\delta g_{\sigma\lambda}
\ee
At this point, we just need to use Eq.~\eqref{K} and note that
\be
u^\sigma
=
\frac{\partial x^\sigma}{\partial X^0}\,U^0
=
c\, \frac{\partial x^\sigma}{\partial X^0}
\left[
g_{\mu\nu}\,\frac{\partial x^\mu}{\partial X^0}\,\frac{\partial x^\nu}{\partial X^0}
\right]^{-1/2}
\ ,
\ee
in order to finally obtain
\be
\delta\left(\sqrt{-g}\,\rho\right)
=
\frac{1}{2}\,\sqrt{-g}
\left(\rho\, c^{-2}u^\mu\,u^\nu
\right)
\delta g_{\mu\nu}
=
-
\frac{1}{2}\,\sqrt{-g}
\left(\rho\,c^{-2}u_\mu\,u_\nu
\right)
\delta g^{\mu\nu}
\ .
\ee
{Expanding the variation in the left-hand-side of this expression and using} the well-known relation for the variation of the metric determinant, we obtain
\be \label{varrho}
\delta\rho
=
\frac{\rho}{2}
\left(g_{\mu\nu}
-
c^{-2} u_\mu\,u_\nu
\right)
\delta g^{\mu\nu}
\ .
\ee
For a barotropic perfect fluid, the pressure is a function of the proper energy density only, $p=p(\epsilon)$. In this case the baryonic density $\rho$ and the proper energy density $\epsilon$ are related
to each other by means of the following relation~\cite{Chavanis:2014lra}
\be
\frac{\delta\rho}{\rho}
=
\frac{\delta\epsilon}{\epsilon+p(\epsilon)} \ .
\ee
Combining this relation with Eq.~\eqref{varrho}, finally yields the variation~\eqref{deps} of the proper energy density relevant for the derivation of the field equations in the Markov-Mukhanov formalism reported in Sect.~\ref{sec:fieldEq}.

\bibliographystyle{unsrt}
\bibliography{bibtov}

\end{document}